\documentclass[twocolumn,showpacs,a4paper,pre]{revtex4}
\usepackage{graphicx}
\usepackage{amsmath}
\usepackage{amssymb}

\begin{document}

\title{Star-graph expansions for bond-diluted Potts models}
\date{\today}

\author{Meik Hellmund}
\email{Meik.Hellmund@itp.uni-leipzig.de}
\homepage{http://www.physik.uni-leipzig.de/~hellmund}
\author{Wolfhard Janke}
\email{Wolfhard.Janke@itp.uni-leipzig.de}
\affiliation{Institut f{\"u}r Theoretische Physik, Universit{\"a}t Leipzig,
Augustusplatz 10/11, D-04109 Leipzig, Germany}

\begin{abstract}
We derive high-temperature series expansions for the free energy and the
susceptibility of random-bond $q$-state Potts models on hypercubic lattices
using a star-graph expansion technique. This method enables the exact
calculation of quenched disorder averages for arbitrary uncorrelated coupling
distributions. Moreover, we can keep the disorder strength $p$ as well as the
dimension $d$ as symbolic parameters. By applying several series analysis
techniques to the new series expansions,
one can scan large regions of
the $(p,d)$ parameter space for any value of $q$. For the bond-diluted
 4-state
Potts model in three dimensions, which exhibits a rather
strong first-order phase transition in the undiluted case, we present
 results for the transition 
temperature and the effective critical exponent $\gamma$
as a function of $p$ as obtained from the analysis of
susceptibility series up to order 18. A comparison with recent Monte Carlo
data (Chatelain {\em et al.}, Phys.~Rev.~E64, 036120(2001)) 
 shows signals for the softening to a second-order transition
at finite disorder strength.
\end{abstract}

\maketitle

\section{Introduction}

Systematic series expansions for statistical models defined on a lattice are a
well-known method to study phase transitions and critical phenomena
\cite{domb3}. They provide an useful complement to large-scale numerical
simulations, in particular for quenched, disordered systems where the
average over many different disorder realizations is numerically very
time consuming and only some points in the vast parameter space of the systems
can be sampled with realistic effort.

Using high-temperature series expansions, on the other hand, one can obtain
for many quantities results which are
 exact up to a certain order in the inverse
temperature. Here the infinite-volume limit can be taken without problems and
the quenched disorder is treated exactly.
Moreover, one can keep the disorder strength $p$ as well as the
dimension $d$ as symbolic parameters and therefore analyse much larger regions
of the parameter space of disordered systems.
 To this end we developed further the method of
``star-graph expansion'' which allows us to take the disorder average on
the level of individual graphs exactly and apply it to $q$-state 
Potts models with a
bimodal quenched distribution of ferromagnetic couplings.
Using optimized cluster algorithms for the exact calculation of spin-spin
correlators on graphs with arbitrary inhomogeneous couplings,
we obtained series up to order 18 in the inverse temperature
for the susceptibility and the free energy of bond-diluted
Potts models in two, three and four dimensions.  

Depending on the dimension $d$ and the number of states $q$, pure Potts
models show first- or second-order phase transitions. According to the Harris
criterion~\cite{harris}
one expects in the second-order case either the appearance of a new
random fixed point ($d=2$, $q=3,4$ and $d=3$, $q=2$) or
logarithmic corrections
to the pure fixed point ($d=2$, $q=2$).
At first-order transitions, randomness softens the transitions. For $d=2$ even
infinitesimal disorder induces a continuous transition~\cite{aiz}, whereas for
$d=3$, $q>2$ a tricritical point at a finite disorder strength is
expected~\cite{card97}.
This  softening to a second-order phase transition
beyond a tricritical point at some finite disorder strength has recently been
verified in Monte Carlo (MC) simulations
of the three-dimensional {\em site\/}-diluted 3-state~\cite{balles00}
and {\em bond\/}-diluted 4-state~\cite{chat01a} Potts model.

The critical part of the series expansion methods
 lies in the extrapolation techniques which are used
in order to obtain information on the critical singularity from
a finite number of known coefficients of the high-temperature series.
One can question the use of these extrapolation techniques in disordered
systems, where the complete singularity structure of the function may be very
complicated, involving Griffiths-type singularities or logarithmic
corrections~\cite{card99}.

Anyhow, we  are able to determine  the transition temperature
for the bond-diluted 4-state Potts model in three dimensions
reliably up to the vicinity of the geometrical percolation point and in good
agreement with analytic estimates~\cite{turb} and MC
results~\cite{chat01a}.

The critical exponent $\gamma$ extracted from our analysis appears to be dependent
on the disorder strength which is caused by  crossover effects and the
complicated singularity structure. Using  sophisticated analysis methods,
we find a range of the disorder strength where 
$\gamma_{\rm eff}=1$, indicating critical behavior
governed by a tricritical point.

The rest of the paper is organized as follows.
In Sect.~\ref{sec:m} we briefly recall the model.
In Sect.~\ref{sec:s} we describe the
methods used for generating the series, and Sect.~\ref{sec:a}
is devoted to a representation  of the analysis techniques used
and their application to the study of the  bond-diluted 4-state Potts model
in three dimensions.

\section{Model}
\label{sec:m}
The $q$-state Potts model on the hypercubic lattice $\mathbb{Z}^d$, or more
generally on any
graph $G$ with arbitrary coupling
constants $J_{ij}$ assigned to the links $\langle ij\rangle$
of $G$, is defined by its partition function
\begin{equation}
  \label{eq:1}
  Z = \sum_{\{S_i\}} \exp \left(\beta \sum_{\langle ij\rangle }  J_{ij} \delta(S_i, S_j)\right),
\end{equation}
where $\beta=1/k_B T$ is the inverse temperature, $S_i=1,\ldots,q$ and
$\delta(.,.)$ is the Kronecker symbol.
In our series expansion the combination
\begin{equation}
  \label{eq:v2}
 v_{ij} = \frac{e^{\beta J_{ij}}-1}{e^{\beta J_{ij}}-1+q}
\end{equation}
will be the relevant expansion parameter.
In the symmetric high-temperature phase, the
susceptibility corresponding to the coupling to an external field $h_i$,
$\sum_i h_i \left(\frac{q \delta(S_i,1) -1}{q-1}\right)$,
is given for a  graph with $N$ spins  by
\begin{equation}
  \label{eq:s1}
  \chi = \frac{1}{N} \sum_i \sum_j \left[ \left\langle \frac{q \delta(S_i, S_j)
        -1}{q-1}\right\rangle \right].
\end{equation}
Quenched disorder averages $[\ldots]$ are taken
over an uncorrelated bimodal distribution of the form
\begin{equation}
  \label{eq:bi}
  P(J_{ij}) = (1-p) \delta(J_{ij}-J_0) + p \delta(J_{ij}- RJ_0),
\end{equation}
which can include spin glasses ($R=-1, p=1/2$),
random-bond ferromagnets ($0<R<1$) and
bond dilution ($R=0$)
as special cases. Other distributions can,
in principle, also be considered with our method.

\section{Series generation}
\label{sec:s}

\subsection{Basic notations from graph theory}
A  graph of order $E$  consists of $E$ links connecting $N$ vertices.
We consider only connected, undirected  graphs that are simple:
 no link starts and ends at the
same vertex and two vertices are never connected by more than one link.
Subgraphs are defined by the deletion of links. In this process, isolated
vertices can be dropped. A graph of order $E$ has $2^E$
(not necessarily non-isomorphic) subgraphs
 since each link may be present or absent.
These subgraphs may consist of several  connected components
and are called clusters.

An articulation point is a vertex the deletion of which renders the graph
disconnected. A graph without articulation points is called ``star graph''.

A graph is bipartite if the vertices can be separated into red and black
vertices so that no link connects two vertices of the same
color. Equivalently, all closed paths in the graph consist of an even number
of links.

\subsection{Star-graph expansion method}

There are two well-established methods \cite{domb3} for the systematic
generation of high-temperature series expansions,
the linked cluster and the star-graph
method. The longest known series (up to order $\beta^{25}$)
for classical spin models without disorder
are produced by linked cluster expansions \cite{butera02}. This technique
allows one to obtain series for observables (such as the second moment of the
spin-spin correlation function) which have no star-graph
expansion. Furthermore, it works with free embeddings of graphs into the
lattice which can be counted orders of magnitude faster than the weak
embedding numbers used by the star-graph technique. Nonetheless, this method
has not yet been applied to problems with quenched disorder.

The star-graph method can be adopted to systems involving quenched disorder
\cite{rap1,singh87} since it allows one to take the disorder average
on the level of individual graphs.
The basic idea is to assemble the  value of some extensive thermodynamic
quantity $F$ on a large or even infinite graph from its values on subgraphs:
Graphs constitute a partially ordered set under the ``subgraph'' relation.
Therefore, for every function  $F(G)$ defined on the set of graphs
exists another function $W_F(G)$ such that for all graphs $G$
\begin{equation}
  \label{eq:2}
   F(G) = \sum_{g \subseteq  G} W_F(g),
\end{equation}
and this function can be calculated recursively via
\begin{equation}
  \label{eq:3}
  W_F(G) = F(G) -  \sum_{g \subset   G} W_F(g).
\end{equation}
This gives for an infinite (e.g. hypercubic) lattice
\begin{equation}
  \label{eq:4}
  F(\mathbb{Z}^d) = \sum_G (G:\mathbb{Z}^d)\, W_F(G),
\end{equation}
where $(G:\mathbb{Z}^d)$ denotes the weak
embedding number of the graph $G$ in the given
lattice structure~\cite{martin74}.

The following observation makes this a useful method:
Let $G$ be  a  graph with an articulation vertex
where two star subgraphs $G_{1,2}$ are glued together.
Then $W_F(G)$ vanishes if
\begin{equation}
  \label{eq:sg}
F(G) = F(G_1) + F(G_2).
\end{equation}
An observable $F$ for which Eq.~(\ref{eq:sg}) is true on arbitrary graphs
with articulation points allows a star-graph expansion. All non-star graphs
have zero weight $W_F$ in the sum Eq.~(\ref{eq:4}).

It is easy to see that the (properly normalized) free energy $\log Z$
has this property and it can be proved~\cite{singh87} that the inverse
susceptibility $1/\chi$ has it, too, even for arbitrary inhomogeneous couplings
$J_{ij}$. This restricts the sum in Eq.~(\ref{eq:4}) to a sum over star graphs.
The linearity of Eqs.~(\ref{eq:2})-(\ref{eq:4}) enables the calculation of
quenched averages over the coupling distribution on the level of individual
graphs.

The resulting recipe for the susceptibility series is:

\begin{itemize}
\item Graph generation and  embedding number counting.
\item Calculation of $Z(G)$ and the correlation matrix\\
$M_{nm}(G) = \mathrm{Tr}\, (q\delta(S_n,S_m)-1) e^{-\beta H(\{J_{ij}\})}$\\
for all graphs as polynomials in $E$ variables $v_{ij}$.
\item Inversion of the $Z$ polynomial as a series up to the desired order.
\item Averaging over quenched disorder,\\
$N_{nm}(G) = \left[ M_{nm}/Z \right]_{P(J)},$\\
  resulting in a matrix of polynomials in $(p,v)$.
\item Inversion of the matrix $N_{nm}$  and subgraph subtraction,\\
 $W_\chi (G) = \sum_{n,m} (N^{-1})_{nm} - \sum_{g\subset G} W_\chi(g)$.
\item Collecting the results from all graphs,\\
$1/\chi = \sum_G (G: \mathbb{Z}^d)\; W_\chi(G)$.
\end{itemize}

\subsection{Generation of star graphs and calculation of embedding numbers}

The most complicated part in every attempt to
generate lists of graphs  by recursively adding nodes and edges
to a smaller list
is the isomorphism test, i.e., the decision whether two
given  adjacency lists or adjacency matrices describe the same graph modulo
relabelling and reordering of edges and nodes. We used the \texttt{nauty}
package by
McKay \cite{mckay81} which makes very fast  isomorphism tests by
calculating a canonical representation of the
automorphism group of the graphs.

{\squeezetable
\begin{table}
\caption{\label{tab:1}Number of star graphs with $E$ links and
non-vanishing embedding numbers on $\mathbb{Z}^d$.}
\begin{ruledtabular}
  \begin{tabular}{l|*{15}{r|}r}
    order $E$&1 &4&6&7&8&9&10&11&12&13&14&15&16&17&18&19\\
    \hline
    $\#$  &1&1&1&1&2&3&8&9&29&51&142&330&951&2561&7688&23078\\
  \end{tabular}
\end{ruledtabular}
\end{table}
}

Since we are only interested in star graphs with non-vanishing
weak embedding numbers in $\mathbb{Z}^d$,  the following simple observations are
helpful:
\begin{itemize}
\item Only bipartite graphs occur since $\mathbb{Z}^d$  is bipartite.
\item A generic $k$-dimensional embedding  (i.e. one which really needs all
   $k$ dimensions)  contributes in $d>k$
  dimensions with degeneracy  $\binom{d}{k}$.
\item A biconnected graph of odd order $E=2n+1$ has generic embeddings only up
  to dimension $n$ since it must have at least two edges in each dimension.
\item The only biconnected graph of even order $E=2n$ which has generic
  embeddings of dimension $n$ is the cycle of length $2n$. All the other
  graphs will use at most $n-1$ dimensions.
\end{itemize}

For the embedding count we implemented 
 a refined version of the backtracing algorithm by
Martin \cite{martin74}.  We did extensive tests to find the optimal algorithm
for the ``innermost'' loop, the test for collisions in the embedding, and ended
up using optimized hash tables.

\begin{figure}
\unitlength6mm
\begin{center}
  \begin{picture}(8,1)
\thicklines
\put(0,0){\line(1,0){8}}
\put(0,0){\line(0,1){1}}
\put(0,1){\line(1,0){8}}
\put(8,0){\line(0,1){1}}
\put(3,0){\line(0,1){1}}
\put(0,0){\circle*{.24}}
\put(1,0){\circle*{.24}}
\put(2,0){\circle*{.24}}
\put(3,0){\circle*{.24}}
\put(4,0){\circle*{.24}}
\put(5,0){\circle*{.24}}
\put(6,0){\circle*{.24}}
\put(7,0){\circle*{.24}}
\put(8,0){\circle*{.24}}
\put(0,1){\circle*{.24}}
\put(1,1){\circle*{.24}}
\put(2,1){\circle*{.24}}
\put(3,1){\circle*{.24}}
\put(4,1){\circle*{.24}}
\put(5,1){\circle*{.24}}
\put(6,1){\circle*{.24}}
\put(7,1){\circle*{.24}}
\put(8,1){\circle*{.24}}
\end{picture}
\end{center}
\small
\begin{eqnarray*}
   & 7620 \binom{d}{2} +  76851600    \binom{d}{3}+ 14650620864 \binom{d}{4}\\
 &+\; 404500471680\binom{d}{5}+ 3355519311360   \binom{d}{6}
\end{eqnarray*}

\begin{center}
  \begin{picture}(6,3)
\thicklines
\put(0,0){\line(1,0){3.9}}
\put(1,1){\line(1,0){3.9}}
\put(0,0){\line(1,1){1}}
\put(3.9,0){\line(1,1){1}}
\put(4.9,1){\line(0,1){1.3}}
\put(2.6,0){\line(0,1){1.3}}
\put(1.3,1.3){\line(1,0){2.6}}
\put(2.3,2.3){\line(1,0){2.6}}
\put(1.3,1.3){\line(1,1){1}}
\put(2.6,1.3){\line(1,1){1}}
\put(3.9,1.3){\line(1,1){1}}
\put(0,0){\circle*{.24}}
\put(1,1){\circle*{.24}}
\put(1.3,0){\circle*{.24}}
\put(2.6,0){\circle*{.24}}
\put(3.9,0){\circle*{.24}}
\put(2.3,1){\circle*{.24}}
\put(3.6,1){\circle*{.24}}
\put(4.9,1){\circle*{.24}}
\put(1.3,1.3){\circle*{.24}}
\put(2.6,1.3){\circle*{.24}}
\put(3.9,1.3){\circle*{.24}}
\put(2.3,2.3){\circle*{.24}}
\put(3.6,2.3){\circle*{.24}}
\put(4.9,2.3){\circle*{.24}}
\end{picture}
\end{center}
\small
\begin{eqnarray*}
   &12048 \binom{d}{3}+  396672\binom{d}{4} +  2127360\binom{d}{5}+
2488320\binom{d}{6}
\end{eqnarray*}

  \caption{Two star graphs of order 17 and 19
and their weak embedding numbers up to 6 dimensions.}
  \label{fig:emb}
\end{figure}

By this means, we
classified for the first time all star graphs up to order 19 which can be
embedded in hypercubic lattices (see Table~\ref{tab:1}) and
calculated their (weak) embedding numbers for $d$-dimensional hypercubic
lattices (up to order 17 for arbitrary $d$,  order 18 and 19 for
dimensions $\leq4$), see Fig.~\ref{fig:emb} for  typical results.

\subsection{Cluster representation}

The partition function and the matrix  of
correlations $M_{nm}$ for each graph are calculated
with arbitrary symbolic couplings $J_{ij}$ using the cluster representation
\begin{eqnarray}
  \label{eq:cl1}
  Z &=& q^{N-E} \prod_{\langle ij\rangle} (e^{\beta J_{ij}}-1+q) \; {\mathcal Z},  \\
  {\mathcal Z} &=& q^{-N} \mathrm{Tr} \prod_{\langle ij\rangle}\left[1-v_{ij} +v_{ij} q \delta(S_i,S_j)
    \right]
\label {eq:cl3}\\
   &=&\!\sum_{C}  q^{e+c-N}\!\! \left(\prod_{\langle ij\rangle \in C}
 v_{ij}\!\right)\!\!\left( \prod_{\langle ij\rangle \notin C} (1-v_{ij})\! \right)\!.\label{eq:cl2}
\end{eqnarray}
Here  the sum goes
over all clusters $C \subseteq G$, $E$ is the number of links ($=$ order) of the
graph $G$, $e$  is the number of links of
the cluster and $c$ the number of connected components of $C$.
${\mathcal Z}$ is normalized such that $\log \mathcal Z$ has a star-graph
expansion.  This essentially reduces the partition sum from a sum over $q^N$
states to a sum over $2^E$ clusters.
In the Ising case $q=2$ another huge
simplification takes place
 since only clusters where all vertices are of even degree contribute
to the cluster sum.

The $2^E$ clusters belonging to a graph
are enumerated by Gray codes~\cite{numrec}
such that two consecutive clusters in the sum (\ref{eq:cl2}) differ by exactly
one (added or deleted) link. Gray codes are a reordering of the binary
representation of numbers such that the difference to the successor is in
exactly one bit position. For example, for $E=4$  the sequence is
0000, 0001, 0011, 0010, 0110, 0111, 0101, 0100, 1100, 1101,
1111, 1110, 1010, 1011, 1001, 1000 where zeros denote the deleted links.
This allows to speed up the calculation
considerably by re-using every term in the sum for the calculation of the next
one.

The calculation of the susceptibility involves the matrix of correlations
$M_{nm}$. The effect of inserting
$\frac{q \delta(S_i, S_j) - 1}{q-1}$ into the trace of
Eq.~(\ref{eq:cl3})  can easily be seen: we get one if the vertices $n$ and $m$
 belong to the same connected component of the cluster and zero otherwise.
Therefore,
\begin{equation}
  \label{eq:clu}
M_{nm} \propto  \sum_{C_{nm}} q^{e+c-N} \left( \prod_{\langle ij\rangle \in C}
 v_{ij} \right)\left( \prod_{\langle ij\rangle \notin C} (1-v_{ij})\right),
\end{equation}
where the sum is restricted to all  clusters $C_{nm}\subseteq G$ in
which the vertices $n$ and $m$ are connected.

For the symbolic calculations
we developed a C\raise2pt\hbox{\tiny++}
template library using an expanded degree-sparse
representation of polynomials and series in many variables.
 The open source library GMP is used  for the
arbitrary-precision arithmetics.

Our longest series, up to order 18, are obtained for the case of bond dilution
where (\ref{eq:bi}) simplifies to
\begin{equation}
  \label{eq:dil}
    P(J_{ij}) = (1-p) \delta(J_{ij}-J_0) + p \delta(J_{ij}),
\end{equation}
since in this case the disorder average for a series is most easily done
via
\begin{equation}
  \label{eq:bd}
 [v_1^{n_1}\ldots v_k^{n_k}]_{P(J)} = (1-p)^k v_0^{n_1+\ldots+n_k}.
\end{equation}

\section{Series analysis: techniques and results}
\label{sec:a}

In the following we shall illustrate the analysis using the
  bond-diluted 4-state Potts model
in three dimensions as our primary  example.
This model
 exhibits in the pure case a strong
first-order transition which is expected to stay first order up to some
finite disorder strength, before it gets softened to a second-order transition
governed by a disordered fixed point.

In the latter case we
 are interested in locating power-law divergences in the susceptibility
series of the form
\begin{equation}
  \label{eq:div}
  \chi (v) = A (v_c-v)^{-\gamma} + \ldots
\end{equation}
For such a critical behavior 
many different series analysis techniques have been
discussed in the literature 
which all have their merits and drawbacks~\cite{guttmann89}.

To localize a first-order transition point, however,
a high-temperature series alone is not sufficient since there  
 the correlation length remains finite and no critical singularity occurs. 
In analysing series by ratio, Pad{\'e} or differential approximants, the
approximant will
provide an analytic continuation  of the thermodynamic quantities beyond the
transition point into a metastable region on a pseudo-spinodal line
with a singularity $T^*_c < T_c$ and effective ``critical exponents''
at $T^*_c$.

Employing the techniques described above 
we obtained the high-temperature series expansions for the susceptibility up
to order~18
with coefficients given as polynomials in the disorder strength $p$,
as listed in the Appendix for general dimensions $d\leq4$. For
such a series in two variables, the method of partial differential
approximants~\cite{pda80}
 should be well suited. Up to date, however, the only application
of this method to a tricritical point \cite{adler97} used a test series of
order 50 generated from an exactly solvable model. In our case, it was unable
to give conclusive results. Therefore, we confined ourselves to the analysis of
single-parameter series for selected values of $p$.

\begin{figure}
\includegraphics[scale=.33,angle=-90]{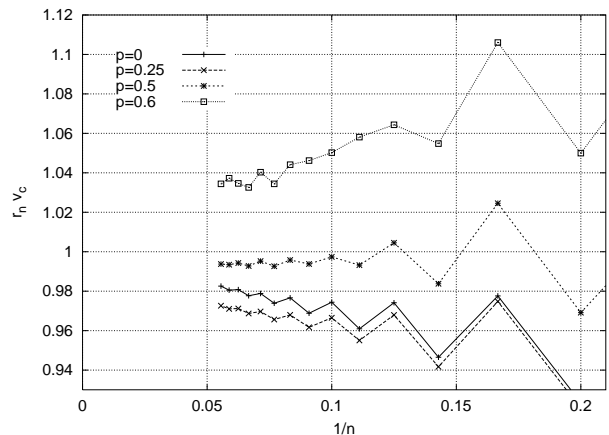}
   \caption{Ratio approximants for different dilutions $p$ vs.\ $1/n$.}
    \label{fig:1}
\end{figure}
The ratio method is the least sophisticated method of series analysis,
but usually it is quite robust and
 gives  a good first estimate of the series behavior.
It  assumes that the expected singularity of the form~(\ref{eq:div})
is the closest to the origin. Then the
consecutive ratios of series coefficients behave asymptotically as
\begin{equation}
  \label{eq:rat}
  r_n = \frac{a_n}{a_{n-1}} = v_c^{-1} \left(1+\frac{\gamma-1}{n}\right).
\end{equation}
Figure~\ref{fig:1} shows these ratios for different values of $p$.
In order to
make them visually comparable, they are normalized by their respective
critical couplings $v_c$.
For small $p$ they show the typical oscillations related to the existence
of an antiferromagnetic singularity at $-v_c$. Near the percolation threshold
at $p=0.751\,188$~\cite{lorenz98}
(where $T_c$ goes to 0) the series is clearly ill-behaved,
related to the $\exp(1/T)$ singularity expected there. Besides that, 
we observe that the slope
($\propto \gamma-1$) is increasing with $p$, changing from $\gamma<1$ to
$\gamma>1$ around $p=0.5$.

The widely used DLog-Pad{\'e} method consists in calculating Pad{\'e}
approximants to the logarithmic derivative of $\chi(v)$. The smallest real
pole of the approximant is an estimation of $v_c$ and its residue gives
$\gamma$.
\begin{figure}
\includegraphics[scale=.33,angle=-90]{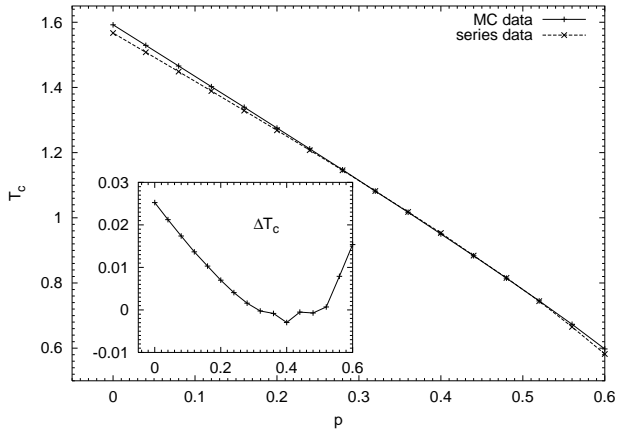}
    \caption{Critical temperature for different dilution $p$ as obtained from
    MC simulations \cite{chat01a} and DLog-Pad{\'e} series analyses.
    The inset shows the difference between the two estimates.}
    \label{fig:2}
\end{figure}
Figure~\ref{fig:2} compares the critical temperature, estimated from an
average of 25-30 Pad{\'e} approximants for each value of $p$~\footnote{%
Notice that ``$p$'' in the present notation corresponds to ``$1-p$'' in
Ref.~\cite{chat01a}.},
with the results  of recent MC simulations \cite{chat01a}.
For small $p$, in the first-order region, the series underestimates the
critical temperature. As explained above, this is an estimate not of $T_c$
but of $T^*_c$. Between $p=0.3$ and $p=0.5$, the estimates confirm,
within errors, the MC results, indicating that now both methods see
the same second-order transition. Beyond $p=0.5$, the scatter of different
Pad{\'e} approximants increases rapidly, related to the crossover to the
percolation point.

\begin{figure}
\includegraphics[scale=.33,angle=-90]{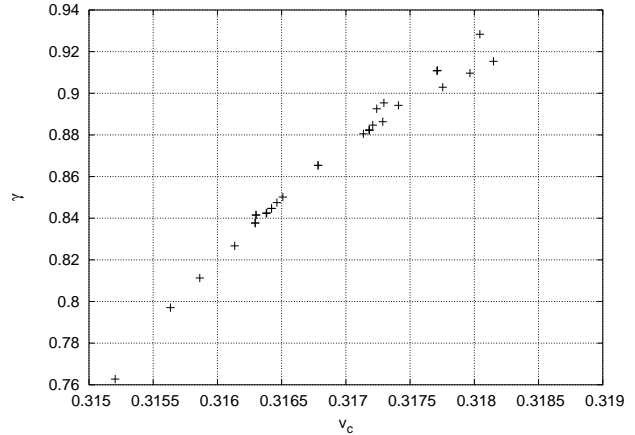}
    \caption{Scattering of different Pad{\'e} approximants at a dilution $p=0.4$:
critical exponent $\gamma$ against critical coupling $v_c$.}
    \label{fig:3}
\end{figure}

\begin{figure}
\includegraphics[scale=.33,angle=-90]{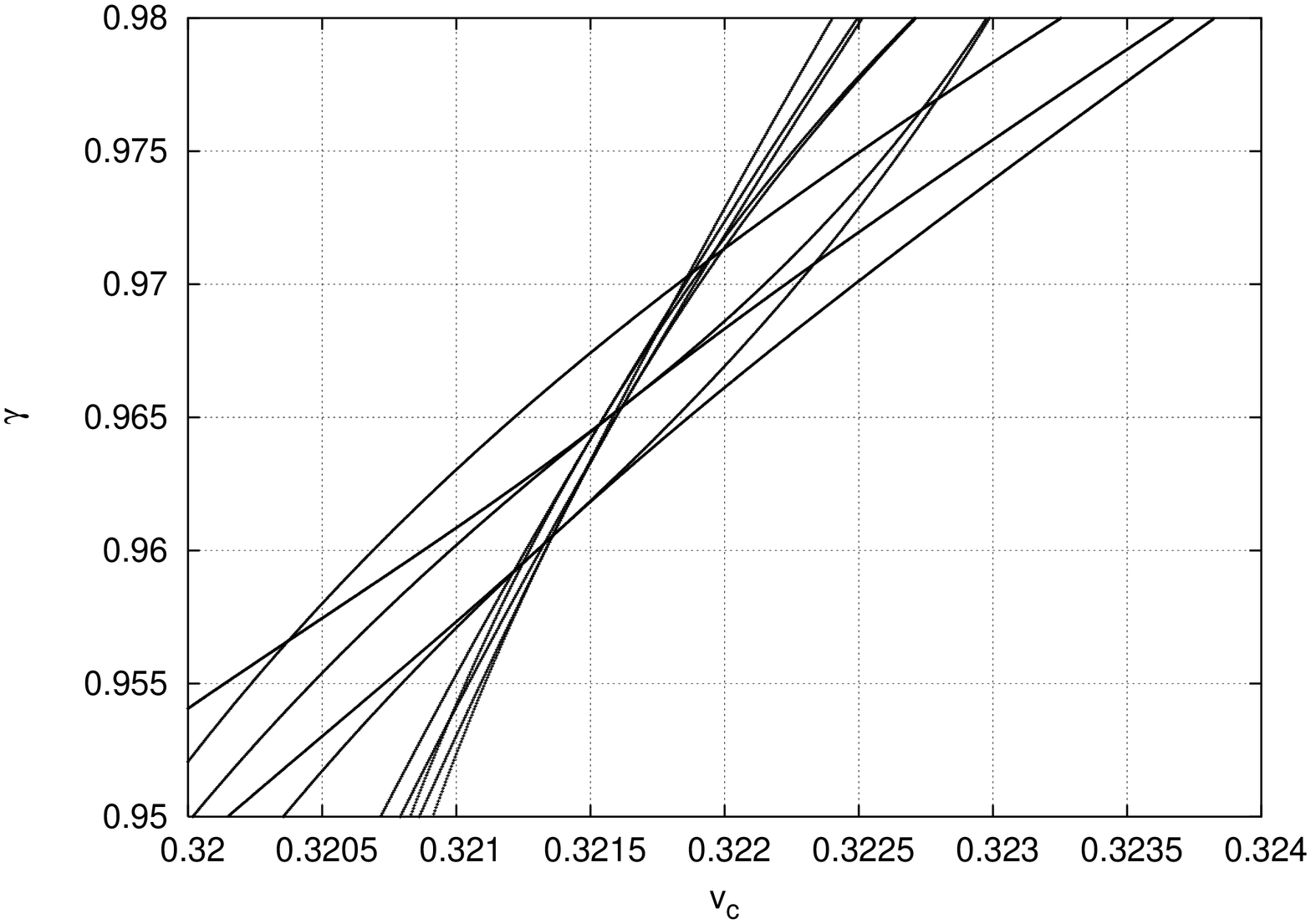}
\includegraphics[scale=.33,angle=-90]{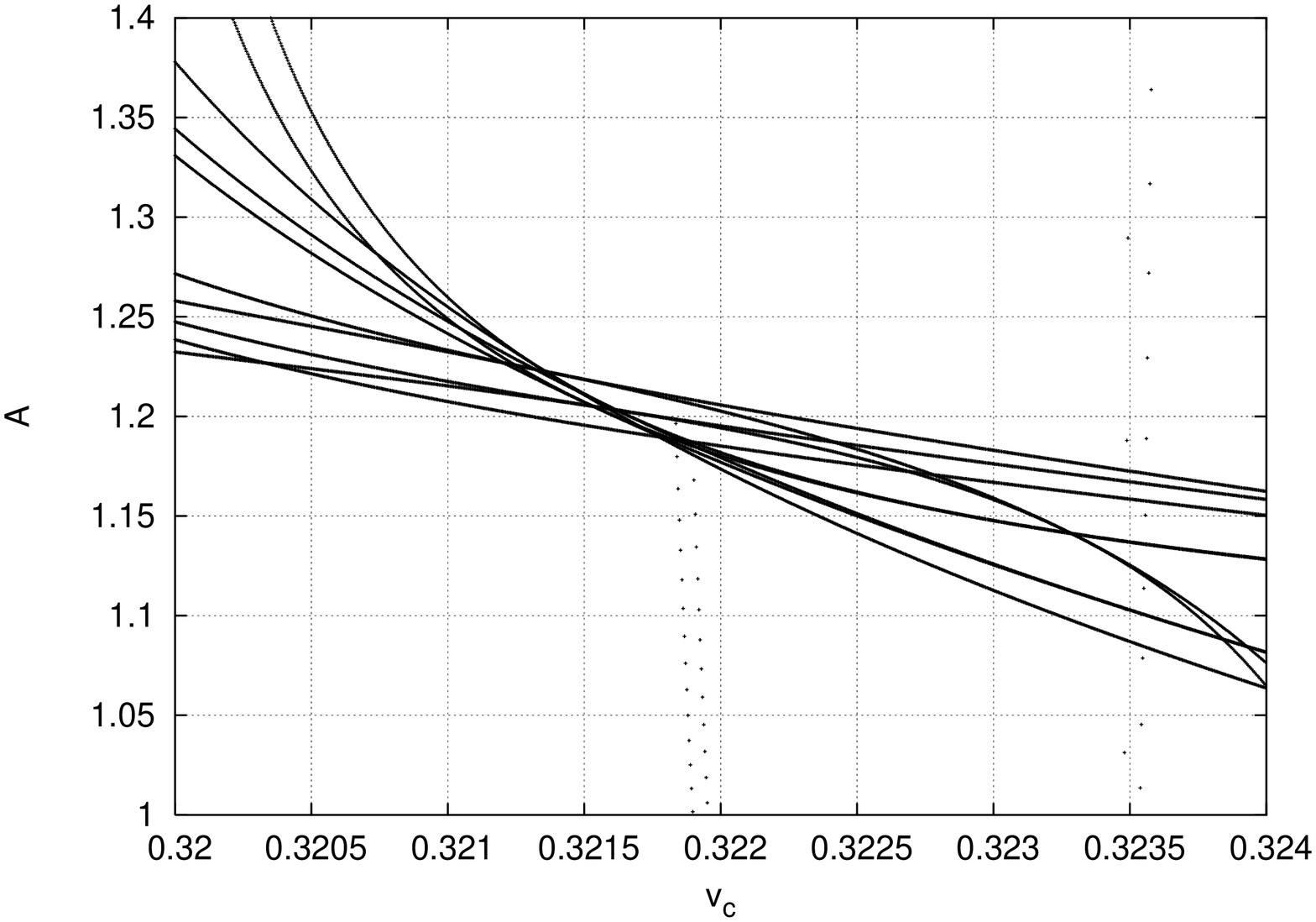}

    \caption{Values for the critical exponent $\gamma$ and  amplitude $A$
at $p=0.4$ as function of trial $v_c$ estimates from the Baker-Hunter analysis.
From the clustering of different Pad{\'e} approximants in both pictures
we estimate  $v_c=0.3217,$ $\gamma=0.966,$ and $A = 1.21$.}
    \label{fig:4}
\end{figure}

The situation is more complicated with respect to the critical exponent $\gamma$.
A DLog-Pad{\'e} analysis gives inconclusive results due to a large scattering
between different Pad{\'e} approximants, as shown in Fig.~\ref{fig:3}. One
possible reason for this failure is the existence of confluent singularities:
The dots in Eq.~(\ref{eq:div}) indicate correction terms which can be
parametrized as follows:
\begin{equation}
  \label{eq:div2}
  \chi (v) = A (v_c-v)^{-\gamma} [1 + A_1(v_c-v)^{\Delta_1} + A_2 (v_c-v)^{\Delta_2} + \ldots],
\end{equation}
where $\Delta_i$ are the confluent correction exponents.
We used  different
more sophisticated analysis methods,
such as inhomogeneous differential approximants~\cite{guttmann89}
and the methods M1 and M2~\cite{adler91},
 especially tailored to deal with
 confluent singularities. In the case at hand,
the Baker-Hunter method~\cite{bakerhunter} appeared to be quite successful,
giving consistent results at larger dilutions $p>0.35$ where the DLog-Pad{\'e}
analysis failed.
 Assume the function under investigation has confluent singularities
 \begin{equation}
   \label{eq:b1}
   F(z) = \sum_{i=1}^N A_i \left(1-\frac{z}{z_c}\right)^{-\lambda_i} = \sum_{n=0} a_n z^n .
 \end{equation}
This can be transformed into an auxiliary function $g(t)$ which is
meromorphic and therefore suitable for Pad{\'e} approximation. After the
substitution $z = z_c(1-e^{-t})$ we expand $F(z(t)) = \sum_n c_n t^n $
and construct the  new series
\begin{equation}
  \label{eq:b2}
g(t) = \sum_{n=0} n!\;c_n\; t^n = \sum_{i=1}^N \frac{A_i}{1-\lambda_it}.
\end{equation}
We see that Pad{\'e} approximants to $g(t)$ have poles at $t=1/\lambda_i$, with
residues at the poles of $-A_i/\lambda_i$.
This method is applied by plotting these poles and residues for different
Pad{\'e} approximants to $g(t)$ as functions of $z_c$. The optimal set of values
for the parameters is determined visually from the best clustering of
different Pad{\'e} approximants, as demonstrated in Fig.~\ref{fig:4}.

\begin{figure}
\includegraphics[scale=.33,angle=-90]{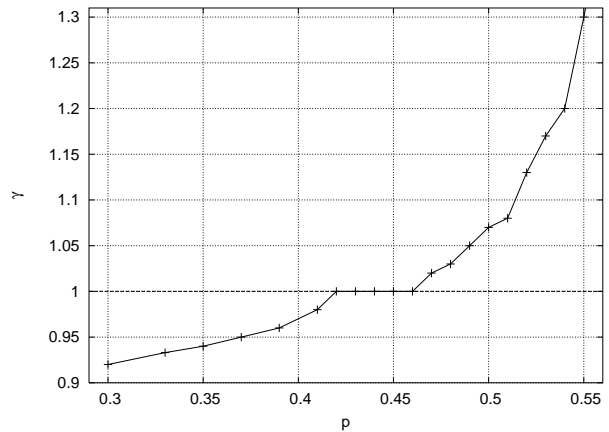}

\caption{Effective critical exponent  $\gamma$ as function of the dilution $p$
from Baker-Hunter analyses.}
  \label{fig:6}
\end{figure}

Using this method, our results for the critical exponent $\gamma$ are plotted in
Fig.~\ref{fig:6}. They show an effective exponent monotonically increasing
with $p$ but reaching a
plateau at $\gamma=1$ for dilutions between
$p=0.42$ and $p=0.46$. The following sharp increase is to be interpreted as
due to the crossover to the percolation fixed point $p_c=0.751\,188$, $T_c=0$,
where a $\chi \sim \exp(1/T)$ behavior is expected.

It is well known (see, e.g., Ref.~\cite{barma85}) 
that series analysis in crossover
situations is extremely difficult. If the parameter $p$ interpolates between 
regions governed by different fixed points, the  exponent obtained from a
 finite number of terms of a series expansion must cross somehow 
between its  universal values, and does this usually quite slowly. 
Therefore it does not come as a
 surprise that the MC simulations quoted above see 
the onset of a second order phase transition already for smaller values of 
the disorder strength $p$.     
 The mere existence of a plateau in $\gamma_{\rm eff}(p)$, however,  
is an indication that here  truly critical behavior is seen. 
It is  governed by a 
fixed point for which we obtain  $\gamma=1.00(3)$. 
Here, as always in series analyses, the error estimates the
scattering of different approximants. 

\section{Conclusions}
\label{sec:c}

We have implemented a comprehensive toolbox for generating and enumerating
star graphs as required for high-temperature series expansions of quenched,
disordered systems.
Monte Carlo simulations of systems with quenched disorder require an enormous
amount of computing time because many realizations have to be simulated for
the quenched average. For this reason it is hardly possible to scan a whole
parameter range. Using high-temperature series expansions, on the other hand,
one can obtain this average exactly. Since the relevant parameters (degree
of disorder $p$, spatial dimension $d$, number of states $q$, etc.)
can be kept
as symbolic variables, the number of potential applications is very large.

Here we  presented an analysis of the three-dimensional bond-diluted 4-state
Potts model. The phase diagram confirms recent Monte Carlo results and, by
comparing with the numerical data, we also see signals for the onset of a
second-order transition at a finite disorder strength.

\begin{acknowledgments}
    Support by  DFG grant No.~JA 483/17-1 and partial support from the
  German-Israel-Foundation under
          grant No.~I-653-181.14/1999 is gratefully acknowledged.
\end{acknowledgments}

\newpage
\onecolumngrid

\appendix*
\section{}
As an example, we publish here the inverse
susceptibility for the bond-diluted 4-state Potts model in $d\leq4 $
dimensions up to order $v^{18}$ ($P=1-p$): \\

\small

$\chi^{-1}(P,v,d) =
 1 - 2\,P\,v\,{d} + 2\,P^2\,v^2\,{d} -
  2\,P^3\,v^3\,{d} +
   \left[ 2\,P^4\,{d} - 16\,P^4\,{d\choose 2} \right]\,v^4   +
  \left[ -2\,P^5\,{d} + \left(24\,P^4 +
     72\,P^5\right){d\choose 2} \right]\,v^5  \\[.2cm] +
   \left[ 2\,P^6\,{d} + \left(24\,P^4 -
     96\,P^5 - 248\,P^6\right){d\choose 2} - 768\,P^6\,{d\choose 3}
 \right]\,v^6 \\[.2cm] +
  \left[ -2\,P^7\,{d} + \left( 24\,P^4 -
     96\,P^5 + 264\,P^6 +
     640\,P^7\right) {d\choose 2} + \left(576\,P^6 +
     3264\,P^7\right){d\choose 3} \right]\,v^7 \\[.2cm] +
  \left[ 2\,P^8\,{d} + \left(72\,P^4 -
     96\,P^5 + 264\,P^6-
     216\,P^7 - 1384\,P^8\right) {d\choose 2} +
     \left(576\,P^6 - 144\,P^7 -
     22704\,P^8\right) {d\choose 3} 
  - 62208\,P^8\,{d\choose 4} \right]\,v^8 \\[.2cm] +
  \left[ -2\,P^9\,{d} + \left(-72\,P^4 -
     288\,P^5 + 264\,P^6 -
     312\,P^7 - 1416\,P^8 +
     1888\,P^9\right) {d\choose 2} \right.\\ \left.\;\;\;  + \left(576\,P^6  -
     720\,P^7  + 720\,P^8 +
     66944\,P^9\right) {d\choose 3} + \left(31104\,P^8 +
     221312\,P^9\right) {d\choose 4} \right]\,v^9 \\[.2cm] +
  \left[ 2\,P^{10}\,{d} +\left(- 72\,P^4 +
     288\,P^5 + 648\,P^6 +
     384\,P^7 - 144\,P^8 +
     9336\,P^9 - 296\,P^{10}\right) {d\choose 2} \right.\\ \left.\;\;\;+
     \left(576\,P^6  + 3456\,P^7 +
     19296\,P^8 + 75456\,P^9 -
     387168\,P^{10}\right){d\choose 3} + \left(31104\,P^8 +
     109440\,P^9 - 4000512\,P^{10}\right){d\choose 4} \right]\,v^{10}
   \\[.2cm] +
  \left[ -2\,P^{11}\,{d} +\left(- 72\,P^4 +
     288\,P^5 - 504\,P^6 -
     1008\,P^7 - 3024\,P^8 +
     3144\,P^9 - 33336\,P^{10} -
     9616\,P^{11}\right){d\choose 2} \right.\\ \left. \;\;\; + \left(576\,P^6 -
     1440\,P^7 - 8352\,P^8 -
     31248\,P^9 - 309744\,P^{10} +
     781824\,P^{11}\right){d\choose 3} \right.\\ \left.\;\;\;
    + \left( 31104\,P^8 +
     11520\,P^9 + 635520\,P^{10} +
     10415872\,P^{11}\right){d\choose 4} \right] \, v^{11} \\[.2cm] +
  \left[ 2\,P^{12}\,{d} +\left(- 216\,P^4 +
     288\,P^5 - 216\,P^6 -
     4272\,P^7 + 240\,P^8 +
     11856\,P^9 - 4968\,P^{10}  +
     81744\,P^{11} + 37320\,P^{12}\right){d\choose 2} \right.\\ \left.\;\;\; +
     \left(2880\,P^6 - 31392\,P^7 +
     14112\,P^8 + 169200\,P^9 +
     489024\,P^{10}
     + 2692800\,P^{11} -
     5811664\,P^{12}\right){d\choose 3} \right.\\ \left. \;\;\;
+ \left(31104\,P^8 +
     273024\,P^9 + 3204864\,P^{10} +
     16037760\,P^{11}   - 179275648\,P^{12}\right){d\choose 4} 
\right] \, v^{12}\\[.2cm] +
 \left[ -2\,P^{13}\,{d} +
     \left(216\,P^4 + 864\,P^5 -
     792\,P^6 - 3912\,P^7 +
     29736\,P^8 + 5952\,P^9 -
     20736\,P^{10} + 23088\,P^{11} \right.\right. \\ \left.\left.\;\;\; -
     144624\,P^{12} - 96160\,P^{13}\right){d\choose 2} \right. \\ \left. \;\;\; +
     \left(-1728\,P^6 - 31536\,P^7 +
     162288\,P^8 + 15408\,P^9 -
     223344\,P^{10} - 113760\,P^{11} -
     8412192\,P^{12} + 5990784\,P^{13}\right){d\choose 3} \right. \\
  \left. \;\;\; + 
     \left(31104\,P^8 + 67968\,P^9 +
     1022976\,P^{10} - 693504\,P^{11} -
     16255872\,P^{12} + 304010112\,P^{13}\right){d\choose 4} 
\right]\,v^{13} \\[.2cm] +
  \left[ 2\,P^{14}\,{d} +
     \left(216\,P^4 - 864\,P^5 -
     1944\,P^6 - 8616\,P^7 +
     18360\,P^8 - 99600\,P^9 -
     65544\,P^{10} + 33936\,P^{11} \right.\right.  \\ \left.   \left. \;\;\; -
     86952\,P^{12} + 73704\,P^{13} +
     169400\,P^{14}\right){d\choose 2} \right. \\ \left. \;\;\;
    + \left( - 1728\,P^6 -
     55152\,P^7 + 67248\,P^8  -
     894240\,P^9 - 918000\,P^{10} +
     2799648\,P^{11} + 8589744\,P^{12} \right.\right. \\ \left.\left.\;\;\; +
     58983984\,P^{13}  - 98045424\,P^{14}\right){d\choose 3}
\right.\\ \left.\;\;\;  +
     \left(31104\,P^8 - 1057536\,P^9 +
     17280\,P^{10} + 24870528\,P^{11} +
     179980416\,P^{12} + 1095494784\,P^{13}  -
     7487817088\,P^{14}\right) {d\choose 4} \right]\, v^{14}\\[.2cm] +
 \left[ -2\,P^{15}\,{d} + \left(216\,P^4 -
     864\,P^5 + 1512\,P^6 -
     4536\,P^7 + 44568\,P^8 -
     55200\,P^9 + 168480\,P^{10}  +
     363072\,P^{11}  - 11832\,P^{12} \right.\right.  \\ \left.   \left. \;\;\;
     +    530040\,P^{13} + 501600\,P^{14} -
     145632\,P^{15}\right){d\choose 2} \right.\\ \left.\;\;\; 
+ \left(- 1728\,P^6 -
     41040\,P^7  + 209232\,P^8 -
     609984\,P^9 + 1319328\,P^{10} +
     7874208\,P^{11} + 5670048\,P^{12} +
     24319296\,P^{13} \right.\right.  \\ \left.   \left. \;\;\; 
   - 141840288\,P^{14} -
     14817536\,P^{15}\right) {d\choose 3} \right.\\ \left.\;\;\; + 
\left(31104\,P^8 -
     740736\,P^9 + 131328\,P^{10} +
     22334976\,P^{11} + 66366720\,P^{12} +
     188319744\,P^{13} \right.\right.  \\ \left.   \left. \;\;\;
- 1467511296\,P^{14} +
     5362518016\,P^{15}\right){d\choose 4} \right]\, v^{15}\\[.2cm] +
  \left[ 2\,P^{16}\,{d} + \left(648\,P^4 -
     864\,P^5 + 1512\,P^6 +
     24336\,P^7 + 38496\,P^8 -
     148008\,P^9 + 1656\,P^{10}   +
     117024\,P^{11} - 1325376\,P^{12} \right.\right.  \\ \left.
   \left. \;\;\; +
     7200\,P^{13} - 1644000\,P^{14} -
     2926176\,P^{15} - 373984\,P^{16}\right) {d\choose 2} 
\right.\\ \left.\;\;\; + \left(
- 1728\,P^6 + 163296\,P^7 +
     105984\,P^8 - 2305728\,P^9 -
     1857888\,P^{10} - 6620544\,P^{11} -
     48148992\,P^{12}  + 18163728\,P^{13} \right.\right.  \\ \left.
   \left. \;\;\;
+  118520640\,P^{14} + 1144225008\,P^{15} -
     1918717248\,P^{16}\right){d\choose 3} \right.\\ \left.\;\;\;
+ \left(217728\,P^8 -
     3438720\,P^9 - 7119360\,P^{10} -
     64137600\,P^{11} - 149601024\,P^{12} +
     1152714240\,P^{13} + 8368094208\,P^{14} \right.\right.  \\ \left.  
 \left. \;\;\;+
     58294742400\,P^{15} - 317165909504\,P^{16}\right) {d\choose 4} 
\right] \, v^{16} \\[.2cm] +
 \left[ -2\,P^{17}\,{d} + \left(-
     648\,P^4 - 2592\,P^5 +
     1512\,P^6 + 15408\,P^7 -
     161712\,P^8 - 123384\,P^9 +  66792\,P^{10}
 + 39264\,P^{11} \right.\right. \\ \left.\left.\;\;\; -
     1976760\,P^{12} + 3413424\,P^{13} +
     848256\,P^{14} + 4241568\,P^{15}  +
     8541960\,P^{16}  + 2398960\,P^{17}\right){d\choose 2} \right. \\ 
  \left. \;\;\; + \left( 
     - 1728\,P^6 + 109728\,P^7 -
     984096\,P^8 - 1584432\,P^9 +
     413424\,P^{10}  - 11287296\,P^{11} -
     32069376\,P^{12} + 206240976\,P^{13}  \right.\right. \\ \left.\left.\;\;\; +
     380730960\,P^{14} + 1087235856\,P^{15}  -
     1859056704\,P^{16} - 2643006384\,P^{17}\right){d\choose 3} 
 \right. \\ \left. \;\;\; + \left(
    - 93312\,P^8 - 2522880\,P^9 -
     4468608\,P^{10} - 63930240\,P^{11} -
     127255680\,P^{12} + 1693209600\,P^{13} +
     6161021568\,P^{14} \right.\right. \\ \left.\left.\;\;\; + 23385824256\,P^{15} -
     50368269312\,P^{16} - 105383991680\,P^{17}\right) {d\choose 4} 
\right]\, v^{17}\\[.2cm] +
  \left[ 2\,P^{18}\,{d} + \left(- 648\,P^4 +
     2592\,P^5 + 4104\,P^6 +
     83520\,P^7 - 82080\,P^8 +
     465984\,P^9 + 586080\,P^{10}  +
     605064\,P^{11} - 166248\,P^{12}   \right.\right. \\ \left.\left.\;\;\; +
     8121312\,P^{13} - 4714536\,P^{14} -
     2886168\,P^{15} - 3604536\,P^{16}  -
     20651832\,P^{17}- 7297424\,P^{18}\right){d\choose 2} 
\right.\\ \left.\;\;\; + \left(
   -8640\,P^6 + 518400\,P^7 -
     492480\,P^8 + 3752592\,P^9  +
     12513744\,P^{10} - 23522256\,P^{11} -
     66704640\,P^{12} - 53106912\,P^{13}  \right.\right. \\ \left.\left.\;\;\;
     -  884626272\,P^{14} - 122168448\,P^{15} +
     2205877392\,P^{16} + 22700601216\,P^{17}  -
     42014019168\,P^{18}\right){d\choose 3} \right.\\ \left.\;\;\; 
 + \left(- 93312\,P^8 +
     657024\,P^9 + 23180544\,P^{10} -
     169350912\,P^{11} - 762268032\,P^{12} -
     3977024256\,P^{13} - 10126195200\,P^{14} \right.\right. \\
 \left.\left.\;\;\;
+     47139877632\,P^{15} +
     379559824128\,P^{16} +
     2866361546496\,P^{17} +
     3747410465664\,P^{18}\right){d\choose 4} \right]\, v^{18}
$
\twocolumngrid

\begin{thebibliography}{21}
\expandafter\ifx\csname natexlab\endcsname\relax\def\natexlab#1{#1}\fi
\expandafter\ifx\csname bibnamefont\endcsname\relax
  \def\bibnamefont#1{#1}\fi
\expandafter\ifx\csname bibfnamefont\endcsname\relax
  \def\bibfnamefont#1{#1}\fi
\expandafter\ifx\csname citenamefont\endcsname\relax
  \def\citenamefont#1{#1}\fi
\expandafter\ifx\csname url\endcsname\relax
  \def\url#1{\texttt{#1}}\fi
\expandafter\ifx\csname urlprefix\endcsname\relax\def\urlprefix{URL }\fi
\providecommand{\bibinfo}[2]{#2}
\providecommand{\eprint}[2][]{\url{#2}}

\bibitem[{\citenamefont{Domb and Green}(1974)}]{domb3}
\bibinfo{editor}{\bibfnamefont{C.}~\bibnamefont{Domb}} \bibnamefont{and}
  \bibinfo{editor}{\bibfnamefont{M.~S.} \bibnamefont{Green}}, eds.,
  \emph{\bibinfo{title}{Series Expansions for Lattice Models}},
  vol.~\bibinfo{volume}{3} of \emph{\bibinfo{series}{Phase Transitions and
  Critical Phenomena}} (\bibinfo{publisher}{Academic Press},
\bibinfo{address}{New York},  \bibinfo{year}{1974}).

\bibitem[{\citenamefont{Harris}(1974)}]{harris}
\bibinfo{author}{\bibfnamefont{A.~B.} \bibnamefont{Harris}},
  \bibinfo{journal}{Journal of Physics C: Solid State Physics}
  \textbf{\bibinfo{volume}{7}}, \bibinfo{pages}{1671} (\bibinfo{year}{1974}).

\bibitem[{\citenamefont{Aizenman and Wehr}(1989)}]{aiz}
\bibinfo{author}{\bibfnamefont{M.}~\bibnamefont{Aizenman}} \bibnamefont{and}
  \bibinfo{author}{\bibfnamefont{J.}~\bibnamefont{Wehr}},
  \bibinfo{journal}{Phys. Rev. Lett.} \textbf{\bibinfo{volume}{62}},
  \bibinfo{pages}{2503} (\bibinfo{year}{1989}).

\bibitem[{\citenamefont{Cardy and Jacobsen}(1997)}]{card97}
\bibinfo{author}{\bibfnamefont{J.}~\bibnamefont{Cardy}} \bibnamefont{and}
  \bibinfo{author}{\bibfnamefont{J.~L.} \bibnamefont{Jacobsen}},
  \bibinfo{journal}{Phys. Rev. Lett.} \textbf{\bibinfo{volume}{79}},
  \bibinfo{pages}{4063} (\bibinfo{year}{1997}).

\bibitem[{\citenamefont{Ballesteros et~al.}(2000)\citenamefont{Ballesteros,
  Fern{\'a}ndez, Mart{\'\i}n-Mayor, Mu{\~n}oz~Sudupe, Parisi, and
  Ruiz-Lorenzo}}]{balles00}
\bibinfo{author}{\bibfnamefont{H.~G.} \bibnamefont{Ballesteros}},
  \bibinfo{author}{\bibfnamefont{L.~A.} \bibnamefont{Fern{\'a}ndez}},
  \bibinfo{author}{\bibfnamefont{V.}~\bibnamefont{Mart{\'\i}n-Mayor}},
  \bibinfo{author}{\bibfnamefont{A.}~\bibnamefont{Mu{\~n}oz~Sudupe}},
  \bibinfo{author}{\bibfnamefont{G.}~\bibnamefont{Parisi}}, \bibnamefont{and}
  \bibinfo{author}{\bibfnamefont{J.~J.} \bibnamefont{Ruiz-Lorenzo}},
  \bibinfo{journal}{Phys. Rev. B} \textbf{\bibinfo{volume}{61}},
  \bibinfo{pages}{3215} (\bibinfo{year}{2000}).

\bibitem[{\citenamefont{Chatelain et~al.}(2001)\citenamefont{Chatelain, Berche,
  Janke, and Berche}}]{chat01a}
\bibinfo{author}{\bibfnamefont{C.}~\bibnamefont{Chatelain}},
  \bibinfo{author}{\bibfnamefont{B.}~\bibnamefont{Berche}},
  \bibinfo{author}{\bibfnamefont{W.}~\bibnamefont{Janke}}, \bibnamefont{and}
  \bibinfo{author}{\bibfnamefont{P.~E.} \bibnamefont{Berche}},
  \bibinfo{journal}{Phys. Rev. E} \textbf{\bibinfo{volume}{64}},
  \bibinfo{pages}{036120} (\bibinfo{year}{2001}).

\bibitem[{\citenamefont{Cardy}(1999)}]{card99}
\bibinfo{author}{\bibfnamefont{J.}~\bibnamefont{Cardy}},
  preprint \eprint{cond-mat/9911024}.

\bibitem[{\citenamefont{Turban}(1980)}]{turb}
\bibinfo{author}{\bibfnamefont{L.}~\bibnamefont{Turban}},
  \bibinfo{journal}{Journal of Physics C: Solid State Physics}
  \textbf{\bibinfo{volume}{13}}, \bibinfo{pages}{L13} (\bibinfo{year}{1980}).

\bibitem[{\citenamefont{Butera and Comi}()}]{butera02}
\bibinfo{author}{\bibfnamefont{P.}~\bibnamefont{Butera}} \bibnamefont{and}
  \bibinfo{author}{\bibfnamefont{M.}~\bibnamefont{Comi}},
  preprint \eprint{hep-lat/0204007}.

\bibitem[{\citenamefont{Singh and Chakravarty}(1987)}]{singh87}
\bibinfo{author}{\bibfnamefont{R.~R.~P.} \bibnamefont{Singh}} \bibnamefont{and}
  \bibinfo{author}{\bibfnamefont{S.}~\bibnamefont{Chakravarty}},
  \bibinfo{journal}{Phys. Rev. B} \textbf{\bibinfo{volume}{36}},
  \bibinfo{pages}{546} (\bibinfo{year}{1987}).

\bibitem[{\citenamefont{Rapaport}(1972{\natexlab{a}})}]{rap1}
\bibinfo{author}{\bibfnamefont{D.~C.} \bibnamefont{Rapaport}},
  \bibinfo{journal}{Journal of Physics C: Solid State Physics}
  \textbf{\bibinfo{volume}{5}}, \bibinfo{pages}{1830, 2813}
  (\bibinfo{year}{1972}{\natexlab{a}}).

\bibitem[{\citenamefont{Martin}(1974)}]{martin74}
\bibinfo{author}{\bibfnamefont{J.~L.} \bibnamefont{Martin}}, in  \cite{domb3},
  pp. \bibinfo{pages}{97--112}.

\bibitem[{\citenamefont{McKay}(1981)}]{mckay81}
\bibinfo{author}{\bibfnamefont{B.~D.} \bibnamefont{McKay}},
  \bibinfo{journal}{Congressus Numerantium} \textbf{\bibinfo{volume}{30}},
  \bibinfo{pages}{45} (\bibinfo{year}{1981}),
  \bibinfo{note}{http://cs.anu.edu.au/{$\sim$}bdm/nauty/}.

\bibitem[{\citenamefont{Press et~al.}(1992)\citenamefont{Press, Teukolsky,
  Vetterling, and Flannery}}]{numrec}
\bibinfo{author}{\bibfnamefont{W.~H.} \bibnamefont{Press}},
  \bibinfo{author}{\bibfnamefont{S.~A.} \bibnamefont{Teukolsky}},
  \bibinfo{author}{\bibfnamefont{W.~T.} \bibnamefont{Vetterling}},
  \bibnamefont{and} \bibinfo{author}{\bibfnamefont{B.~P.}
  \bibnamefont{Flannery}}, \emph{\bibinfo{title}{Numerical Recipes in C}}
  (\bibinfo{publisher}{Cambridge University Press}, 
\bibinfo{address}{Cambridge},
\bibinfo{year}{1992}).

\bibitem[{\citenamefont{Guttmann}(1989)}]{guttmann89}
\bibinfo{author}{\bibfnamefont{A.~J.} \bibnamefont{Guttmann}}, in
 vol.~\bibinfo{volume}{13} of
  \emph{\bibinfo{series}{Phase Transitions and Critical Phenomena}}, 
edited by
\bibinfo{editor}{\bibfnamefont{C.}~\bibnamefont{Domb}} \bibnamefont{and}
  \bibinfo{editor}{\bibfnamefont{J.~L.} \bibnamefont{Lebowitz}}, 
  (\bibinfo{publisher}{Academic Press}, \bibinfo{address}{New York},
  \bibinfo{year}{1989}), pp. \bibinfo{pages}{1--234}.

\bibitem[{\citenamefont{Fisher and Chen}(1982)}]{pda80}
\bibinfo{author}{\bibfnamefont{M.~E.} \bibnamefont{Fisher}} \bibnamefont{and}
  \bibinfo{author}{\bibfnamefont{J.-H.} \bibnamefont{Chen}}, in
  \emph{\bibinfo{booktitle}{Phase Transitions: Carg{\`e}se 1980}}, edited by
  \bibinfo{editor}{\bibfnamefont{M.}~\bibnamefont{L{\'e}vy}},
  \bibinfo{editor}{\bibfnamefont{J.~C.} \bibnamefont{Le~Guillou}},
  \bibnamefont{and}
  \bibinfo{editor}{\bibfnamefont{J.}~\bibnamefont{Zinn-Justin}}
  (\bibinfo{publisher}{Plenum}, \bibinfo{address}{New York},
  \bibinfo{year}{1982}), pp. \bibinfo{pages}{169--216}.

\bibitem[{\citenamefont{Salman and Adler}(1997)}]{adler97}
\bibinfo{author}{\bibfnamefont{Z.}~\bibnamefont{Salman}} \bibnamefont{and}
  \bibinfo{author}{\bibfnamefont{J.}~\bibnamefont{Adler}},
  \bibinfo{journal}{Journal of Physics A: Mathematical and General}
  \textbf{\bibinfo{volume}{30}}, \bibinfo{pages}{1979} (\bibinfo{year}{1997}).

\bibitem[{\citenamefont{Lorenz and Ziff}(1998)}]{lorenz98}
\bibinfo{author}{\bibfnamefont{C.~D.} \bibnamefont{Lorenz}} \bibnamefont{and}
  \bibinfo{author}{\bibfnamefont{R.~M.} \bibnamefont{Ziff}},
  \bibinfo{journal}{Phys. Rev. E} \textbf{\bibinfo{volume}{57}},
  \bibinfo{pages}{230} (\bibinfo{year}{1998}).

\bibitem[{\citenamefont{Adler et~al.}(1991)\citenamefont{Adler, Aharony,
  Harris, Klein, and Meir}}]{adler91}
\bibinfo{author}{\bibfnamefont{L.}~\bibnamefont{Klein}},
\bibinfo{author}{\bibfnamefont{J.}~\bibnamefont{Adler}},
  \bibinfo{author}{\bibfnamefont{A.}~\bibnamefont{Aharony}},
  \bibinfo{author}{\bibfnamefont{A.~B.} \bibnamefont{Harris}},
   \bibnamefont{and}
  \bibinfo{author}{\bibfnamefont{Y.}~\bibnamefont{Meir}},
  \bibinfo{journal}{Phys. Rev. B} \textbf{\bibinfo{volume}{43}},
  \bibinfo{pages}{11249} (\bibinfo{year}{1991}).

\bibitem[{\citenamefont{George A.~Baker and Hunter}(1973)}]{bakerhunter}
\bibinfo{author}{\bibfnamefont{G.~A.}~\bibnamefont{Baker}}
  \bibnamefont{and} \bibinfo{author}{\bibfnamefont{D.~L.}
  \bibnamefont{Hunter}}, \bibinfo{journal}{Phys. Rev. B}
  \textbf{\bibinfo{volume}{7}}, \bibinfo{pages}{3377} (\bibinfo{year}{1973}).

\bibitem[{\citenamefont{Barma and Fisher}(1985)}]{barma85}
\bibinfo{author}{\bibfnamefont{M.}~\bibnamefont{Barma}} \bibnamefont{and}
  \bibinfo{author}{\bibfnamefont{M.~E.} \bibnamefont{Fisher}},
  \bibinfo{journal}{Phys. Rev. B} \textbf{\bibinfo{volume}{31}},
  \bibinfo{pages}{5954} (\bibinfo{year}{1985}).

\end{thebibliography}

\end{document}